\documentclass[]{elsarticle}
\usepackage{amssymb}
\usepackage{amsmath}
\usepackage{mathbbol} 
\usepackage[margin=1in]{geometry}
\graphicspath{{Figures/}}
\usepackage{graphics}
\usepackage{rotating} 
\usepackage[normalem]{ulem}
\usepackage{color}
\usepackage{multirow}

\usepackage[utf8]{inputenc}
\usepackage{hyperref}

\sloppypar

\begin{document}
\begin{frontmatter}
\title{Monitoring Electron Spin Fluctuations with Paramagnetic Relaxation Enhancement}
\author[WIS]{Daniel Jard\'{o}n-\'{A}lvarez} 
\author[WIS]{Tahel Malka}
\author[MagLab]{Johan van Tol}
\author[WIS2]{Yishay Feldman}
\author[WIS2]{Raanan Carmieli}
\author[WIS]{Michal Leskes}
\address[WIS]{Department of Molecular Chemistry and Materials Science, Weizmann Institute of Science, Rehovot, 76100, Israel}
\address[MagLab]{National High Magnetic Field Laboratory, Florida State University, 1800 E. Paul Dirac Dr, Tallahassee, FL 32310, United States}
\address[WIS2]{Department of Chemical Research Support, Weizmann Institute of Science, Rehovot, 76100, Israel}

\begin{abstract}
The magnetic interactions between the spin of an unpaired electron and the surrounding nuclear spins can be exploited to gain structural information, to reduce nuclear relaxation times as well as to create nuclear hyperpolarization via dynamic nuclear polarization (DNP). A central aspect that determines how these interactions manifest from the point of view of NMR is the timescale of the fluctuations of the magnetic moment of the electron spins.  These fluctuations, however, are elusive, particularly when electron relaxation times are short or interactions among electronic spins are strong. Here we map the fluctuations by analyzing the ratio between longitudinal and transverse nuclear relaxation times T$_1$/T$_2$, a quantity which depends uniquely on the rate of the electron fluctuations and the Larmor frequency of the involved nuclei.
 This analysis enables rationalizing the evolution of NMR lineshapes, signal quenching as well as DNP enhancements as a function of the concentration of the paramagnetic species and the temperature, demonstrated here for  LiMg$_{1-x}$Mn$_x$PO$_4$  and  Fe(III) doped  Li$_4$Ti$_5$O$_{12}$, respectively. For the latter, we observe a linear dependence of the DNP enhancement and the electron relaxation time within a temperature range between 100 and 300~K.

\end{abstract}

\end{frontmatter}

\section{Introduction}

Many functional materials require the presence of d- and f-block transition metals for achieving highest performance. Properties of interest, such as electronic and ionic conductivity, optical activity or chemical reactivity can strongly depend on their presence. Relevant concentrations can cover the entire range, from very low, where they act as a perturbing dopant, to very high, constituting a major component of the structure. Understanding the effect of the different components within the structure of the materials on the desired properties is fundamental for the design of new materials. For this purpose, NMR spectroscopy represents an ideal characterization method, as it has the capability of mapping the local structure around nuclei of interest.

Transition metals often have unpaired electrons and are therefore paramagnetic in nature. The presence of paramagnetic agents in a sample has profound consequences on its NMR properties. Through space dipolar couplings and Fermi contact interactions can result in large frequency shifts and broadening as well as reduced relaxation times.\cite{nmrpm,pnmrs_111_1_2018}
In the limit of fast electron relaxation time, the hyperfine couplings result in a shift of the nuclear frequency rather than a splitting. This shift is attributed to the mean magnetic moment over the populated electronic spin states. 
The magnitude of the shift will depend on the strength of the interaction, determined by the distance and orbital overlap between  nuclear and electron spin. Consequently, valuable structural information can be extracted from paramagnetic shifts.
In addition, electronic relaxation represents a fluctuation of the magnetic moment of the electron spin, which in turn can cause nuclear relaxation. In rigid solids, where intrinsic motions are inefficient in causing relaxation, this is likely to become a dominant relaxation mechanism.
Paramagnetic relaxation enhancement (PRE) through addition of small amounts of paramagnetic species is a common strategy used  to increase the sensitivity per time in NMR experiments.\cite{ssnmr_5_151_1995,cr_111_530_2011,cm_19_5742_2007,jmr_333_107097_2021}

Fluctuations of local magnetic fields are most efficient in causing longitudinal relaxation when the rate of fluctuation is in the order of the Larmor frequency. Transverse relaxation is in addition also mediated by slow fluctuations.\cite{pomr}
As a consequence, the observed NMR response can be fundamentally distinct depending on the timescale of the electron spin relaxation, or more generally, of the fluctuations of the electron magnetic moment. Specifically, fluctuations of the longitudinal component of the magnetic moment of the electron spins are responsible for the nuclear relaxation.\cite{pr_166_279_1968} We will denote the correlation time of this fluctuation $\tau_{1e}$ and note that it might differ from the electron relaxation time T$_{1e}$ due to the presence of coherent processes.
When the correlation time is very fast ($\tau_{1e}<10^{-9}\text{ s}$), the resulting nuclear relaxation is in the extreme narrowing regime.\cite{ponm} Thus, transverse relaxation times will be relatively long and, in principle, nuclear spins even in the immediate proximity of the paramagnetic center can be observed.
In this regime NMR spectra are very rich in structural information, which can be extracted from the paramagnetic shifts and line broadening,\cite{pnmrs_111_1_2018,cr_104_4493_2004,jacs_134_17178_2012} even in cases where paramagnetic centers are only introduced as minor dopants.\cite{cm_25_3979_2013,jpcc_120_11111_2016,ac_73_128_2017,cm_29_3538_2017} 
For longer correlation times ($\tau_{1e}>10^{-9}\text{ s}$), the slower fluctuations of the electronic magnetic moment become very efficient in causing transverse nuclear relaxation. This can lead to significant line broadening and ultimately signal quenching.\cite{pccp_21_10185_2019}
While short transverse relaxation times are rarely beneficial for NMR, the presence of slow relaxing electron spins can  actually be exploited for dynamic nuclear polarization (DNP) purposes. This scenario is the basis for metal ions based (MI)DNP.\cite{jacs_133_5648_2011,jacs_136_11716_2014,cpc_19_2139_2018,CIC_MIDNP_2021} The sensitivity gained from DNP can be used to obtain valuable structural information in inorganic solids from nuclear sites otherwise inaccessible via NMR due to low signal intensity.\cite{jacs_141_451_2019,jpcc_124_7082_2020,jpcl_12_2964_2021,jacs_143_4694_2021,jpcc_125_18799_2021}
Finally, one could imagine a case where the electron relaxation rate becomes so slow that the hyperfine splitting will no longer be averaged\cite{jmr_326_106939_2021} and coherence lifetimes can become larger with an appropriate averaging scheme, such as MAS.\cite{jmr_65_252_1985} To our knowledge this  has not been observed for paramagnetic metal centers at relevant temperatures and thus will not  be further considered here.

At low concentrations of paramagnetic species, when couplings among electron spins are very weak, the origin of the random fluctuations of the electronic magnetic moments can be attributed to electron spin relaxation, originating from couplings of the electron spin with the lattice, and $\tau_{1e}\approx\text{T}_{1e}$. At increasing concentrations, exchange and dipolar couplings among electron spins become more prominent and can act as an additional source of fluctuations mediating nuclear relaxation. The presence of strong interactions can lead to coherent and incoherent fluctuations of the electron magnetic moments, since the former are not strictly stochastic $\tau_{1e}$ can become shorter than T$_{1e}$.\cite{amr_49_1235_2018} The effect of coherent processes on NMR relaxation can in theory be removed by an appropriate averaging scheme. Of course, this is not a trivial task, nonetheless, it has been proven possible experimentally.\cite{jpcc_122_1932_2018}
In principle, electron relaxation times can be determined via EPR,\cite{rops,jmr_322_106875_2021} in some cases it is even possible to discern relaxation from coherent processes, like spin and spectral diffusion.\cite{emagres_8_295_2019} However, when electron relaxation times are short, measuring them can be challenging  due to instrumental requirements, namely sufficient power and short dead times following microwave pulses. This often impedes the ability to determine electron relaxation times at conditions of interest for MAS NMR and MAS DNP, at relatively high magnetic fields and temperatures.
An alternative approach to analyze the electron spin properties is studying its effect on surrounding nuclear spins. In NMR relaxation dispersion experiments nuclear relaxation times are analyzed over a large range of magnetic fields from which the relevant correlation times can be obtained from fits.\cite{nmrpm} It is also possible to quantify electron relaxation times from measured Overhauser DNP enhancements, as they are related to saturation efficiencies.\cite{jmr_133_1_1998} { Alaniva et al. measured T$_{1e}$ relaxation times indirectly through quantification of the solid effect DNP enhancement after a variable delay succeeding saturation of the electron spins \cite{ac_58_7259_2019}. This experiment, however,  was performed at 4.3~K and required the use of chirped microwave pulses.}

 In this work we focus on obtaining the electron spin fluctuation time from the ratio of  T$_1$ over T$_2$.\cite{jpcl_8_5871_2017} This ratio is particularly useful because it is independent of the strength of the hyperfine coupling. Therefore, it can be used even in the presence of a large distribution of distances to directly determine the correlation time of the electron spin fluctuations, $\tau_{1e}$. The main requirements are that  relaxation is governed by PRE and outside of the extreme narrowing regime (thus, $\tau_{1e}\omega_L > 1$). The ratio T$_1$/T$_2$ approaches unity within the extreme narrowing regime and consequently becomes independent of the correlation time. 
 We note that this method does not discriminate between different sources of fluctuations of the electron magnetic moment. 
  By computing the  ratio T$_1$/T$_2$, we are able to rationalize the observed features in the paramagnetic NMR response and relate these to changes in nature and coordination environment of the paramagnetic ions, but also in concentration and temperature.
 In particular, we study  changes in lineshape and signal quenching in the olivine type material LiMg$_{1-x}$Mn$_x$PO$_4$ when going from a purely diamagnetic ($x=0$) to a strongly paramagnetic material ($x=1$). This system is of particular interest since from extrapolation of the evolution of signal quenching at low concentrations one would expect a complete absence of signal at high concentrations of paramagnetic species; instead, no quenching is observed at $x=1$. Further, we discuss the differences with an analogous series of LiMg$_{1-x}$Fe$_x$PO$_4$, where no signal quenching is observed at any mole fraction $x$. Finally, we show that the decay in DNP enhancement with increasing temperature in Fe(III) doped Li$_4$Ti$_5$O$_{12}$ depends linearly on  the determined $\tau_{1e}$. 
 This approach represents a simple way to qualitatively predict DNP enhancements from paramagnetic centers, with the advantage of being performed at relevant experimental conditions, without additional instrumentation and without the requirement of any further model or assumption.
 
{ The choice of the studied materials is motivated by their electrochemical relevance. The olivines LiMnPO$_4$ and LiFePO$_4$ and the spinel Li$_4$Ti$_5$O$_{12}$ are known cathode and anode materials for lithium ion batteries, respectively.\cite{jmca_1_3518_2013,jac_882_160774_2021,jmca_3_5750_2015} LiMgPO$_4$ can be applied as a solid state lithium ion conductor, whose ion conductivity largely improves upon doping.\cite{cm_27_2074_2015}  In this work we highlight the opportunities and challenges associated with applying NMR in the presence of paramagnetic ions over a wide concentration range, as well as the exciting possibilities of exploiting the dopants as polarizing agents for DNP, for studying technologically relevant materials.}

\section{Methods}

\subsection{Sample Preparation}
Samples of composition  LiMg$_{1-x}$Mn$_x$PO$_4$ with $x = $ 0, 0.001, 0.005, 0.01, 0.02, 0.05, 0.1, 0.2, 0.4, 0.6, 0.8 and 1 and  LiMg$_{1-x}$Fe$_x$PO$_4$ with $x=$ 0.005, 0.01, 0.05 and 0.3 were prepared via solid state synthesis. Stoichiometric amounts of the precursors Li$_2$CO$_3$ (Strem Chemicals, 99.999\%), MgO (Strem Chemicals, 99.99\%), NH$_4$H$_2$PO$_4$ (Strem Chemicals, 99.998\%) and MnCO$_3$ (Alfa Aeser, 99.9\%) or FeC$_2$O$_4\cdot$2H$_2$O (Aldrich, 99.99\%) were ground, pressed into pellets and subsequently heated to 900$^\circ$C or 800$^\circ$C for 8~h and under a steady flow of N$_2$/H$_2$ (95/5), respectively.
Preparation of Fe(III) doped Li$_4$Ti$_5$O$_{12}$ with nominal concentration of $x=0.01$ was described in a previous work.\cite{jpcc_124_7082_2020}

\subsection{X-ray Powder Diffraction}
Powder X-ray diffraction measurements on LiMg$_{1-x}$Mn$_x$PO$_4$ and LiMg$_{1-x}$Fe$_x$PO$_4$ were performed in reflection geometry using Rigaku (Tokyo, Japan) theta-theta diffractometers: an Ultima III equipped with a sealed copper anode tube operating at 40~kV/40~mA and a TTRAX~III equipped with a rotating copper anode X-ray tube operating at 50~kV/200~mA. A scintillation detector was aligned at the diffracted beam after a bent Graphite monochromator, which was used for X-ray cleaning, effectively removing K$_\beta$.
Phase analysis and lattice parameters estimation were performed using Jade Pro software (Materials Data, Inc.) and ICDD database.

\subsection{EPR Measurements}

Continuous wave (CW) X- (9.4~GHz) and Q-band (34~GHz) EPR measurements were performed on a Bruker ELEXYS E-580 spectrometer. High field J-band ($\sim 240$~GHz) CW and pulsed EPR measurements were performed on a  a quasi-optical spectrometer, as described in a previous work,\cite{rsi_79_064703_2008} in an arrangement without resonating structure. Unlike as described in Ref.\cite{rsi_79_064703_2008}, the 4.2~GHz intermediate frequency signal from the primary mixer is down-converted in a IQ mixing scheme using a phase-stable reference that is generated from the base frequencies of the 240 GHz source and 235.8~GHz reference oscillator.   The typical pulse lengths used for the pulsed experiments are 300~ns. Simulations of EPR spectra were done with the EASYSPIN simulation package for Matlab.\cite{jmr_178_42_2006}

Electron relaxation times were measured at high field at temperatures up to 100~K.  At higher temperatures, short coherence lifetimes impede the acquisition of relaxation times due to the low intensity of the Hahn echo.  Longitudinal relaxation times were obtained with the inversion recovery pulse sequence and transverse relaxation times with the Hahn echo pulse sequence with varying echo delay. Relaxation parameters were obtained after fitting the experimental data to a stretched exponential function:
\begin{equation}
M_z(t)=M_z(\infty)\cdot[1-\exp[-(t/T_{1})^{\beta_1}]],
\label{eq:T1satrec}
\end{equation}
and
\begin{equation}
M_{xy}(t)=M_{xy}(0)\cdot\exp[-(t/T_{2})^{\beta_2}].
\label{eq:T2decay}
\end{equation}
Where T$_1$ and T$_2$ are the longitudinal and transverse  relaxation times, respectively, and $\beta_{1,2}$ the corresponding stretched factors.{ Uncertainties are given as one standard deviation of the fit, while experimental sources of error were not considered. Further, the fit parameter T$_{1,2}$ and $\beta_{1,2}$ are strongly correlated. Therefore, we used the obtained values of the stretched factor to asses the nature of the exponential decay, having either a markedly stretched ($\beta$ approaching $0.5$) or simple exponential ($\beta$ approaching $1$) shape, but avoided drawing conclusions from smaller variations of $\beta$.}

\subsection{NMR Measurements}
NMR measurements were performed on a Bruker 9.4 T Avance-Neo spectrometer equipped with a sweep coil and a 263 GHz gyrotron system. Room temperature measurements were done using a 1.3~mm double resonance probe with samples spinning at 60~kHz. The $\pi/2$ pulse lengths were 2~$\mu$s for  $^6$Li,  and 2.625~$\mu$s for $^7$Li and  $^{31}$P. $^{6,7}$Li spectra were referenced using LiF as secondary reference at -1~ppm\cite{essl_8_a145_2005} and $^{31}$P was referenced to ammonium dihydrogen phosphate at 0.8~ppm.\cite{jpc_98_3108_1994}

DNP measurements were performed using a 3.2 mm double resonance LT-DNP probe spinning at approximately 10~kHz. The $\pi/2$ pulse lengths at 100~K were  4 and 2 $\mu$s for $^6$Li and  $^7$Li, respectively. Temperature in variable temperature measurements was determined from $^{79}$Br T$_1$ relaxation times of KBr.\cite{jmr_196_84_2009} Therefore, powdered KBr was ground together with 1\% Fe(III) doped  Li$_4$Ti$_5$O$_{12}$, with a mass ratio of 1:2, prior to packing the rotor. Pulse lengths in variable temperature measurements were calibrated at each temperature.

 Acquisition of  DNP spectra of the LiMg$_{1-x}$Mn$_x$PO$_4$ samples was done with a single-pulse excitation. All other spectra, as well as saturation recoveries, were acquired with the Hahn echo sequence,\cite{pr_80_580_1950}  using an echo delay equivalent to two rotor periods. For quantitative analysis all measurements were preceded by a saturation pulse train and a recycle delay which ensured a recovery of 95\% of the magnetization. Longitudinal magnetization recovery times were obtained using the saturation recovery sequence.\cite{jcp_55_3604_1971} Transverse polarization decay times were measured either with the Hahn echo sequence with varying echo delays or with the CPMG pulse sequence,\cite{pr_94_630_1954,rsi_29_688_1958} the used delays are given in the tables in the Supporting Information (SI). Longitudinal and transverse relaxation times were obtained from fitting the obtained data to equations~(\ref{eq:T1satrec}( and (\ref{eq:T2decay}). The results are summarized in the SI.  For better comparison, the relaxation times from the variable temperature measurements were obtained with a fixed stretched factor of $\beta_{1,2}=0.6$. Fixing $\beta$ did not significantly alter the quality of the fits (see Tables~S8 and S9).


\section{Results and Discussion}

\subsection{Variable paramagnetic dopant concentration}
\subsubsection{X-ray Powder diffraction and EPR characterization}

\begin{figure}
\begin{center}
\includegraphics[scale=1]{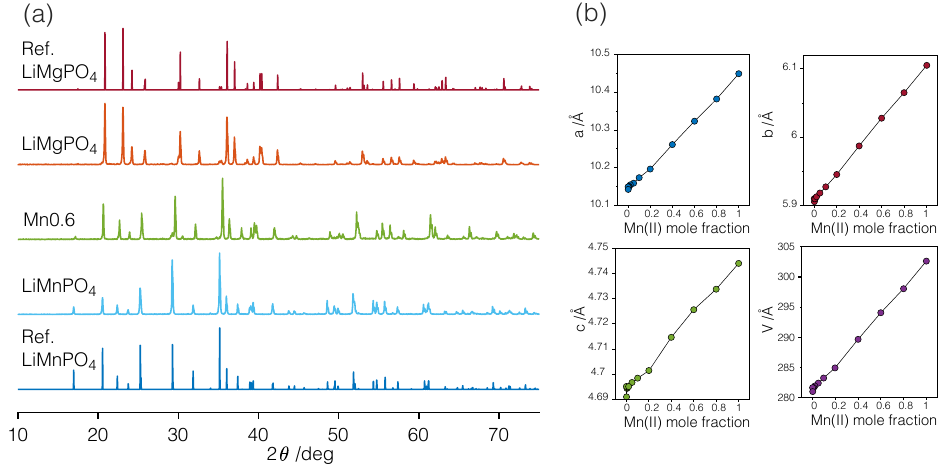}
\end{center}	
\caption{(a) X-ray powder diffraction pattern of  LiMgPO$_4$, LiMg$_{0.4}$Mn$_{0.6}$PO$_4$ and LiMnPO$_4$. (b) Lattice parameters of the orthorombic space group Pnma obtained from refinement of the X-ray diffraction patterns  of the solid solution  LiMg$_{1-x}$Mn$_x$PO$_4$ as a function of the Mn(II) mole fraction. Reference patterns for LiMgPO$_4$ and LiMnPO$_4$ taken from Refs.~\cite{jcsr_12_99_1982} and \cite{ac_13_325_1960}, respectively.}
\label{fg:XRD}
\end{figure}

Fig.~\ref{fg:XRD}a shows the obtained diffraction patterns for selected compositions (diffraction patterns for all compositions are shown in Fig.~S1).  All samples yielded high purity phases. Throughout the entire range of concentration the olivine structure is maintained, indicating that these samples form solid solutions upon exchange of Mg$^{2+}$ with Mn$^{2+}$ or Fe$^{2+}$. In a solid solution the lattice parameters are expected to follow a linear change as a function of the concentration of the substituent in accordance with Vegards law\cite{zfp_5_17_1921}. Analysis of the lattice parameter of  LiMg$_{1-x}$Mn$_x$PO$_4$ over the full range of concentrations from $x=0$ to $x=1$ shows good agreement with the expected behavior as shown in Fig.~\ref{fg:XRD}b. \\

\begin{figure}
\begin{center}
\includegraphics[scale=1]{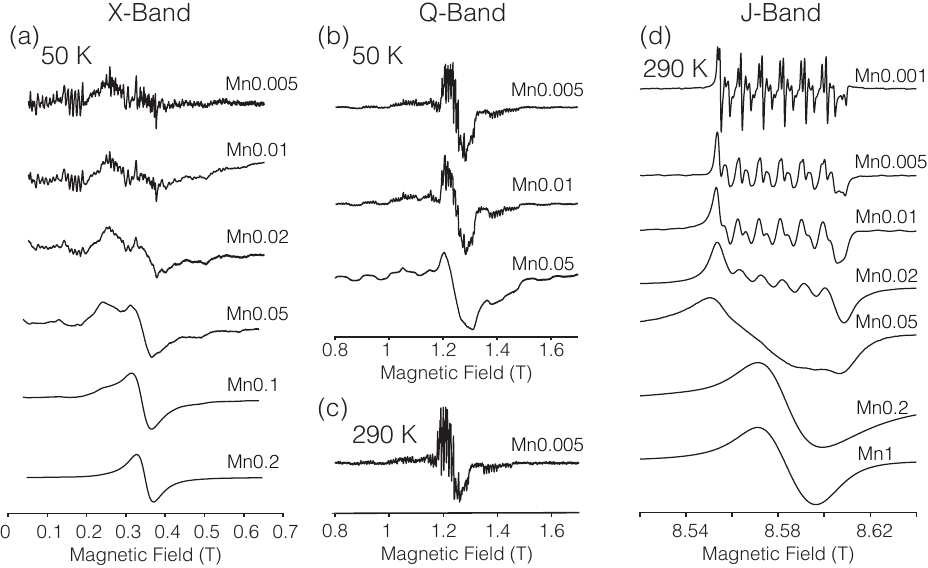}
\end{center}	
\caption{Continuous wave EPR spectra of  LiMg$_{1-x}$Mn$_x$PO$_4$ for selected concentrations of Mn(II) with microwave irradiation frequencies of 9.8 (a), 34.3 (b) and (c) and 240~GHz (d). }
\label{fg:EPRspectra}
\end{figure}

\begin{figure}
\begin{center}
\includegraphics[scale=1]{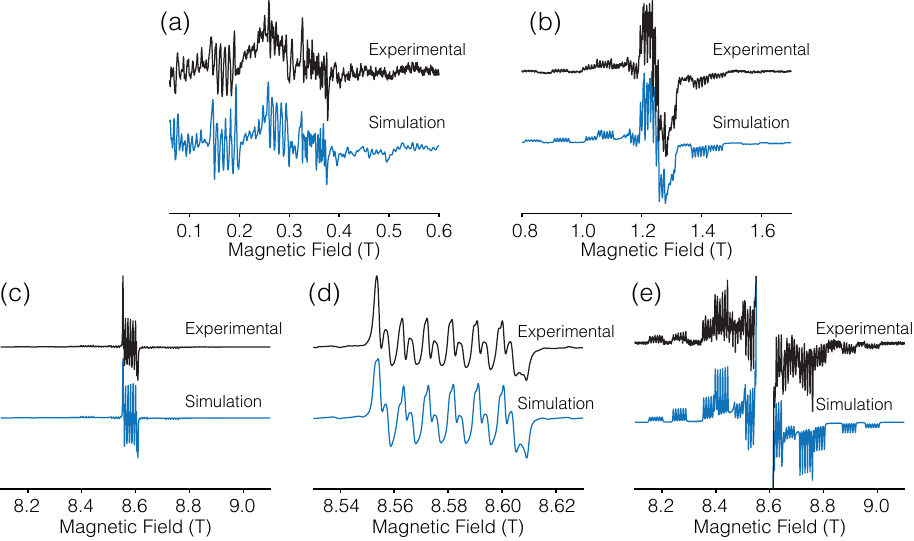}
\end{center}	
\caption{Comparison of experimental data with simulated CW EPR spectra of  LiMg$_{1-x}$Mn$_x$PO$_4$ with $x=0.005$ obtained with best fit parameters $g_{iso}=1.999$, $^{55}$Mn hyperfine coupling $A=259\text{ MHz}$ and zero field splitting parameters $D=2810\text{ MHz}$ and $E=531\text{ MHz}$ with a D-strain of 30~MHz. An additional line broadening was added individually to each of the simulations for best agreement with the experimental results. Simulated and experimental data obtained at microwave irradiation frequencies of 9.8 (a), 34.3 (b) and 240~GHz (c) and temperatures of 50 (a), (b) and 290~K (c). Simulated spectra at 9.8 and 34.4~GHz required  additional shifts for better match experimental results, likely due to small field calibration inaccuracies. Figures (d) and (e) are horizontal and vertical zoom in of figure (c), respectively. All simulations were done with  the EASYSPIN simulation package.\cite{jmr_178_42_2006}}
\label{fg:EPRsim}
\end{figure}

CW EPR spectra of LiMg$_{1-x}$Mn$_x$PO$_4$ for various values of $x$ and at variable fields are shown in  Fig.~\ref{fg:EPRspectra}. Mn(II) has an electronic spin of $S=5/2$ and consequently its EPR spectrum will be affected by the zero field splitting (ZFS) interaction. In addition, the isotope $^{55}$Mn has 100\% natural abundance and a nuclear spin $I=5/2$, hyperfine coupling between the nuclear and electronic spins will lead to a splitting of the EPR line into a characteristic sextet. At low concentrations and fields (Fig.~\ref{fg:EPRspectra}a) the spectra show a complicated lineshape resulting from the combined effect of the hyperfine and zero field splitting (ZFS) interactions. The second order contribution to the ZFS scales inversely with the size of the magnetic field.\cite{emagres_7_179_2018} Consequently, with increasing  field the central transition is less affected by ZFS leading to better resolved spectra (Figs.~\ref{fg:EPRspectra}b and \ref{fg:EPRspectra}d). Simulations reproducing the experimental CW EPR spectra of  LiMg$_{1-x}$Mn$_x$PO$_4$ with $x=0.005$ at X-, Q- and J-band are shown in Fig.~\ref{fg:EPRsim}. Excellent agreement is obtained  using a fixed set of parameters for all microwave irradiation frequencies, and is given in the figure caption.

With increasing concentration of Mn(II) two effects are observed (Fig.~\ref{fg:EPRspectra}). First, at low concentrations, a broadening of the individual peaks occurs. And second, after passing a threshold at approximately $x=0.05$ we observe a narrowing of the overall EPR response. We  attribute the former to an increase of the through space dipolar interaction among electron spins, while the latter to the increasing contribution of the Heisenberg exchange interaction, known to cause line narrowing.\cite{praim}\\

\begin{figure}
\begin{center}
\includegraphics[scale=1]{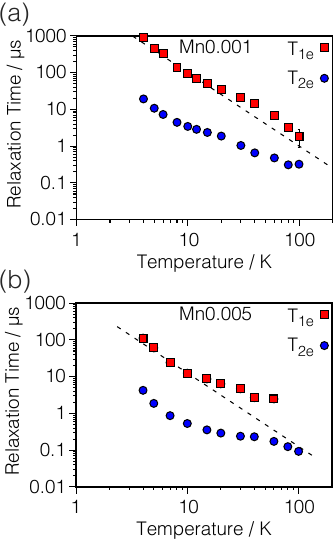}
\end{center}	
\caption{Longitudinal and transverse electron spin relaxation times, T$_{1e}$ and T$_{2e}$, obtained from fit of experimental data to equations~(\ref{eq:T1satrec}) and (\ref{eq:T2decay}) for  LiMg$_{1-x}$Mn$_x$PO$_4$ with $x=0.001$ (a) and $x=0.005$ (b) as a function of temperature. Dashed line shows slope for T$_{1,2}\propto T^{-2}$. Experimental data acquired at a microwave irradiation frequency of 240~GHz.} 
\label{fg:EPRrelax}
\end{figure}

Measured electronic relaxation times T$_{1e}$ and T$_{2e}$ are summarized in Table~S5 and Fig.~\ref{fg:EPRrelax}. The longitudinal relaxation curves show a  stretched exponential behavior with the stretch factor $\beta_1$ ranging around 0.7 for $x=0.001$ and 0.5 for $x=0.005$. The transverse relaxation decay, on the other hand, follows a simple exponential decay. Deviation from exponential behavior of the longitudinal magnetization recovery might be due to contributions from coherent spectral diffusion processes\cite{amr_49_1235_2018}, which become more prominent at higher concentrations, leading to smaller $\beta_1$. The resemblance of the behavior of both relaxation times with temperature, indicate that both are driven by the same relaxation mechanisms. Over the entire measured temperature range T$_{2e}$  is over one order of magnitude shorter than T$_{1e}$ and both decrease steeply with increasing temperature. Above 5~K the relaxation times decrease with a squared temperature dependence (indicated by dashed lines in the figure). This behavior might be attributed to Raman activated relaxation processes.\cite{eprti,jmr_139_165_1999} All curves shown in Fig.~\ref{fg:EPRrelax} show a small region between approximately 10 and 30~K with flatter temperature dependence of the relaxation, after this, the relaxation times return to decay with T$^{-2}$. This weaker temperature dependence is unexpected and is not justified by changes in the stretched factor (see Table~S5), possible explanations for this behavior, could be a change in the contribution from spectral diffusion to the observed relaxation behavior. Alternatively, this effect might be correlated to a shift in the relative contributions of central and satellite transitions with increasing temperature due to the temperature dependence of the Boltzmann distribution. At 100~K T$_{1e}$ in Mn0.001 is 1.9~$\mu$s, it was not possible to measure relaxation times of higher concentrated samples at this temperature. At 60~K T$_{1e}$ in Mn0.005 is 2.5~$\mu$s while at this same temperature T$_{1e}$ in Mn0.001 is 6.8~$\mu s$. 

No EPR signal was observed in the LiMg$_{1-x}$Fe$_x$PO$_4$ samples using X- and Q-band spectrometers and at temperatures ranging from room temperature to 50~K.  The  difference in the EPR behavior between Fe(II) and Mn(II) arises from the different electron configuration. The half-filled (d$^5$) configuration of Mn(II) ensures an isotropic electron density and, therefore, quenching of the spin-orbit coupling.\cite{pnmrs_102_120_2017} Electron relaxation is often mediated by the spin-orbit coupling, thus, absence of significant contribution generally leads to long relaxation times.\cite{nmrpm,acr_40_206_2007} Furthermore, the central transition $+1/2\leftrightarrow-1/2$ is not affected by the zero field splitting to first order. On the other hand, Fe(II) has a d$^6$ electron configuration, and therefore, four unpaired electrons, in the expected high-spin state.\cite{jacs_141_13089_2019} In this case, not only are the orbitals not half-filled, but the even number of electron spins will lead to an integer spin number $S=2$, in this case all transitions will be affected by first order zero field splitting introducing an additional source of line broadening and relaxation. The difficulty of measuring EPR spectra of integer spins is well known and the absence of signal in the measured EPR spectra is a good indication for the absence of Fe(III).

\subsubsection{NMR lineshape and signal quenching}

\begin{figure}
\begin{center}
\includegraphics[scale=1]{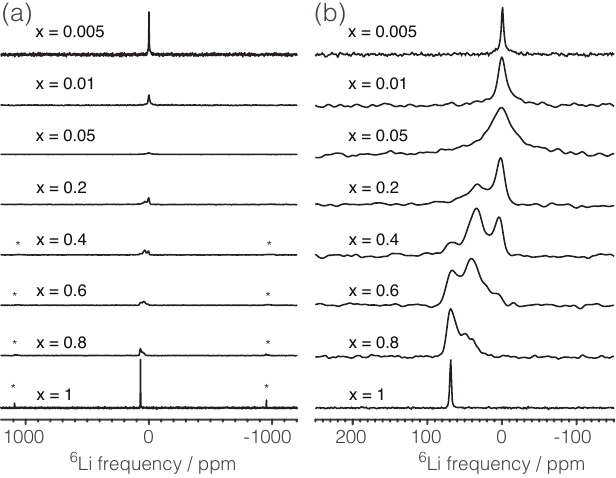}
\end{center}	
\caption{$^6$Li MAS NMR Hahn echo spectra of LiMg$_{1-x}$Mn$_x$PO$_4$ for selected concentrations of Mn(II), obtained at room temperature and a spinning speed $\nu_R$ of 60~kHz. Both figures show the same dataset, in (a) the spectra are normalized according to the number of scans, in (b) spectra are normalized to the maximum intensity.}
\label{fg:Figure6Li}
\end{figure}

\begin{figure}
\begin{center}
\includegraphics[scale=1]{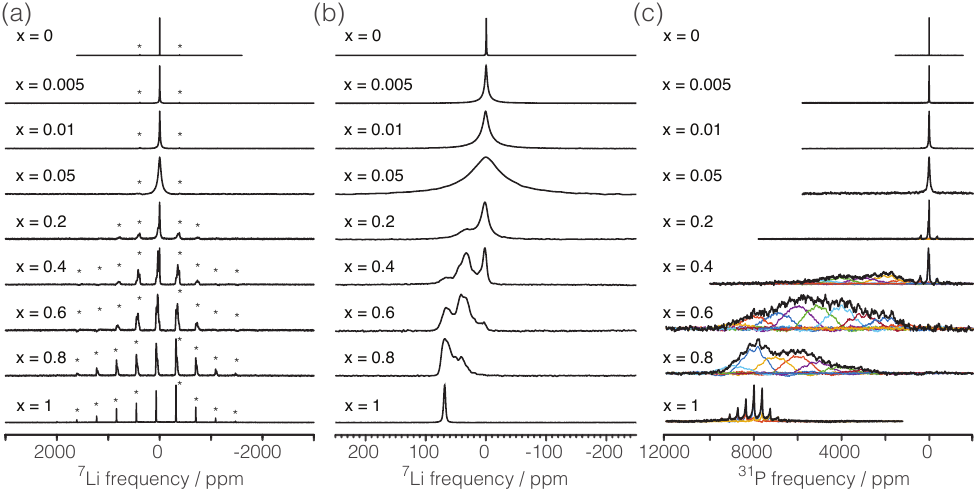}
\end{center}	
\caption{$^7$Li (a) and (b), and $^{31}$P (c) MAS NMR Hahn echo spectra of LiMg$_{1-x}$Mn$_x$PO$_4$ for selected concentrations of Mn(II), obtained at room temperature and a spinning speed $\nu_R$ of 60~kHz. All  spectra are normalized  to the maximum intensity. (a) and (b) show the same dataset. The $^{31}$P spectra (c) with concentrations above $x=0.05$ where obtained from the sum of spectra with different and equally spaced ($1000$~ppm) carrier frequencies, the subspectra are shown in colors.}
\label{fg:Figure7Li31P}
\end{figure}

Figs.~\ref{fg:Figure6Li} and \ref{fg:Figure7Li31P} show the $^{6,7}$Li and $^{31}$P MAS spectra of LiMg$_{1-x}$Mn$_x$PO$_4$ for varying concentration of paramagnetic Mn(II). Although the dopant has only minor effects on the crystal structure of the material, its paramagnetic nature causes drastic changes in the NMR spectrum. The trends are the same for all three nuclei.
In the purely diamagnetic sample the spectra of all nuclei show a single peak, as expected since LiMgPO$_4$ has one unique crystallographic lithium and phosphorous site. The peak is centered at -0.5~ppm for $^6$Li and $^7$Li and at 9~ppm in the case of $^{31}$P. Contributions from chemical shift anisotropy and dipolar couplings are almost entirely removed by spinning at 60~kHz.
Introducing small amounts of paramagnetic dopants causes a broadening of the signal. This homogeneous broadening is attributed to transverse relaxation (vide infra) and reaches a maximum at a Mn(II) concentration of $x=0.05$. 

At high dopant concentrations ($x\geq0.2$) the peaks become sharper again and we observe an increment in the relative intensity of the spinning sidebands as well as the appearance of environments resonating at higher frequencies. The former is attributed to direct through space dipolar couplings while the latter to Fermi contact shifts. 
Further, clearly distinct sites coexisting at a given concentration become visible and can be attributed to differences in the amount and location of Mn(II) in the first coordination shell. In a thorough analysis Cl\'{e}ment et al. showed for the solid solution LiFe$_x$Mn$_{1-x}$PO$_4$ the different contributions of Mn(II) ions to the overall Fermi contact shifts depending on its geometry relative to a given phosphorous nucleus.\cite{jacs_134_17178_2012} In this concentration regime the appearance of the $^{31}$P spectra differs strongly from the $^{6}$Li and $^{7}$Li spectra. The isotropic position in LiMnPO$_4$ is 7996~ppm for $^{31}$P, compared to 68.5~ppm for $^{6,7}$Li.  The markedly stronger covalent character of P--O bonds compared to Li--O bonds results in much higher electron spin density at the $^{31}$P nuclei and consequently larger Fermi contact shifts.  The larger intensity of the spinning sidebands in $^{7}$Li as compared to $^{6}$Li is due to the differences in gyromagnetic ratio of both nuclei. As the isotropic Fermi contact shifts are the same in ppm, the lineshape of the center band is analogous for both nuclei (compare Figs.~\ref{fg:Figure6Li}b and \ref{fg:Figure7Li31P}b). \\

\begin{figure}
\begin{center}
\includegraphics[scale=1]{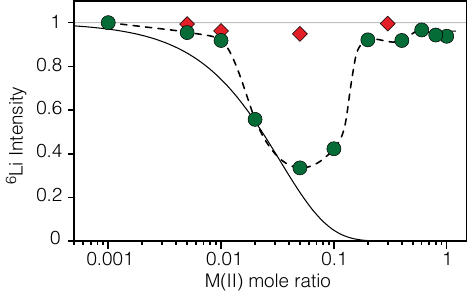}
\end{center}	
\caption{Integrated signal intensity of the $^6$Li Hahn echo spectra of LiMg$_{1-x}$Mn$_x$PO$_4$ (green circles) and LiMg$_{1-x}$Fe$_x$PO$_4$ (red diamonds). { Solid black line shows as guide for the eye an exponential decay of the intensity, which would be expected for a quenching sphere of constant radius around paramagnetic ions, independent of the concentration. }Error bars estimated from signal to noise ratios are smaller than symbols. }
\label{fg:quench}
\end{figure}

The weaker dipolar couplings between electron spin and $^{6}$Li allow excitation of the entire spectrum with a single carrier frequency. Therefore, the $^{6}$Li NMR response was quantified to analyze signal quenching. The signal intensity is shown in Fig.~\ref{fg:quench} for variable Mn(II) concentration. At low Mn(II) contents we observe an  increment of signal quenching with increasing concentration, reaching a minimum signal intensity at a Mn(II) mole fraction of $x=0.05$. At concentrations larger than $x=0.1$ the intensity increases sharply and for $x$ larger than 0.2 again all $^{6}$Li nuclei can be observed. This behavior follows the trend seen in the homogeneous line broadening, as one would expect, since both result from short coherence lifetimes. The initial decrease in intensity is expected and { is known as signal quenching}. The observed behavior can be understood assuming a constant quenching sphere around each paramagnetic species.\cite{pccp_21_10185_2019} Within the radius of this sphere (in this simplified picture we will neglect the angular dependence of the dipolar coupling) the transverse relaxation time of the nuclei is shorter than the echo spacing, thus nuclei within the sphere will not be observed. The decrease in signal intensity with increasing paramagnetic species is then simply a consequence of an increased number of relaxation sources and thus overall quenched volume{, and is expected to follow a simple exponential decay (solid black line in Fig.~\ref{fg:quench})}.\cite{pccp_18_9752_2016,pccp_21_10185_2019} 
 { Similar results have been reported in the context of quantifying sensitivity enhancements\cite{ssnmr_66_6_2015} from DNP using organic radicals\cite{jmr_216_209_2012,cs_3_108_2012,jmr_240_113_2014} as well as metal ions\cite{jmr_240_113_2014,jacs_141_451_2019,jpcc_124_7082_2020,jpcc_125_23126_2020} as polarizing agents.} More interesting, {however,} is the subsequent inversion of the trend. The reduction of quenched signal above a certain concentration threshold, despite an ongoing increment in paramagnetic species, is an indication of a change in the properties of the electronic spin leading to a reduction of the quenching sphere radius. To gain deeper understanding of this observation, in the next section we turn our attention to the paramagnetic enhancement of nuclear relaxation times.

It should be mentioned that one could reduce the quenching factor by direct acquisition after a single excitation pulse instead of acquiring a Hahn echo. However, due to the sharp decay of the signal intensity immediately after excitation, the experimental requirement of a dead time after each pulse will result in a strong baseline distortion making precise quantification of the signal intensity more difficult,  which is the main goal of this analysis.\\

For comparison we measured $^{6,7}$Li and $^{31}$P MAS spectra of the analogous solid solution LiMg$_{1-x}$Fe$_x$PO$_4$ (spectra shown in Fig.~S2). Most notably, we do not observe any signal quenching over the entire concentration range of $x$ (Fig.~\ref{fg:quench}). While we do see broadening of the line with increasing concentration of Fe(II), this line broadening is inhomogeneous in nature\cite{jcp_70_3300_1979} and can be refocused by a $\pi$-pulse. At low concentrations the resonances of the Fe-doped samples are consistently narrower compared to Mn-doped samples at the same concentration. This is in line with the expected effect due to the faster electron relaxation times of Fe(II) and will be discussed further in the next section. 

On the other hand, while at high concentrations of Mn(II) we observe narrowing of the peaks, this does not occur for high Fe(II) concentrations, and thus at values of $x\geq0.2$ the peaks in the Fe-doped samples  are broader than in the equivalent Mn-doped ones, while we do not observe any significant change in the relative intensity of the spinning sidebands. These observations are in agreement with previous results.\cite{jap_84_416_1998,jacs_134_17178_2012} Again, the different effects of Fe(II) and Mn(II) on the NMR spectra arise from the much stronger spin--orbit coupling in the former, which is quenched in the d$^5$ configuration of the latter. Paramagnetic spins with spin--orbit couplings can lead to additional shifts and broadening due to the pseudo-contact interaction and bulk magnetic susceptibility effects.\cite{jacs_141_13089_2019} {As there is no increase in the anisotropic contributions to the spectrum}, the main source of the observed broadening in LiMg$_{1-x}$Fe$_x$PO$_4$ is likely due to the presence of distribution of pseudo contact shifts, while increased structural disorder and anisotropic bulk magnetic susceptibility might have some minor contributions.

\subsubsection{Nuclear relaxation}
 Nuclear relaxation requires fluctuations of local magnetic fields. In rigid solids sources of relaxation can be scarce leading to very long relaxation times, this is particularly pronounced in the absence of strong interactions.
 The presence of paramagnetic impurities will lead to an enhancement in relaxation, as the intrinsically faster relaxation of the electron spin will cause random fluctuations of the local magnetic field of surrounding nuclei. 
 At low concentrations of paramagnetic species, the dominant interaction between electronic spins and most nuclei will be the through space dipolar coupling. The longitudinal relaxation of a nucleus $I$ at a distance $r$ from a paramagnetic center $S$ for this case will be:\cite{pr_166_279_1968}
\begin{equation}
		 \frac{1}{T_{1}}=
\frac{6}{15}\left(\frac{\mu_{0} }{4\pi}\right)^2 
\frac{S(S+1)\gamma_I^2g_e^2\mu_B^2}{r^6}
\left(
\frac{\tau_{1e}}{1+\tau_{1e}^2\omega_I^2}
 \right),
\label{eq:T1PRE}
\end{equation}
transverse relaxation is given by:
\begin{equation}
\frac{1}{T_{2}}= 
\frac{1}{2T_{1}}+
\frac{4}{15}\left(\frac{\mu_{0} }{4\pi}\right)^2 
\frac{S(S+1)\gamma_I^2g_e^2\mu_B^2}{r^6}
\left( 
{\tau_{1e}}
 \right).
\label{eq:T2PRE}
 \end{equation}
Where $\mu_0$ is the vacuum permeability, $S$ the electron spin, $\gamma_I$ the nuclear gyromagnetic ratio, $g_e$ the free electron $g$ value, $\mu_B$ the electron Bohr magneton, $\omega_I$ the nuclear Larmor frequency and $\tau_{1e}$ the correlation time describing the fluctuation of the longitudinal component of the magnetic moment of the electron. We avoid the use of $T_{1e}$ for generality, as coherent processes can contribute to fluctuations of local magnetic fields leading to enhanced nuclear relaxation. Whenever possible we will comment on the origin of the fluctuations. For simplicity, in equations~(\ref{eq:T1PRE}) and (\ref{eq:T2PRE}) we have assumed $\frac{\tau_{2e}}{1+\tau_{2e}^2\omega_S^2}\ll\frac{\tau_{1e}}{1+\tau_{1e}^2\omega_I^2}$, where $\omega_S$ is the Larmor frequency of the electron and $\tau_{2e}$ the correlation time describing the fluctuation of the transverse component of the magnetic moment of the electron. This is a good assumption for any relevant ratio of $\tau_{1e}/\tau_{2e}$.

Equations~(\ref{eq:T1PRE}) and (\ref{eq:T2PRE}) refer to a single electron-nucleus pair. For this case, it has been noted before \cite{jpcl_8_5871_2017} that the ratio between nuclear T$_1$ and T$_2$ from PRE is independent of the strength of the dipolar interaction, and only depends on the correlation time $\tau_{1e}$ and the nuclear Larmor frequency:
\begin{equation}
\frac{T_1}{T_2}=\frac{7}{6}+\frac{4}{6}\tau_{1e}^2\omega_I^2 
\label{eq:T1T2ratio}
\end{equation}
Consequently, one can simply obtain the size of $\tau_{1e}$ from the ratio of T$_1$/T$_2$:
\vspace{0.2cm}
\begin{equation}
 \tau_{1e}=\sqrt{\left(\frac{T_1}{T_2}-\frac{7}{6}\right)\frac{6}{4\omega_I^2}}
\label{eq:T1efromT1T2}
\end{equation}
This equation will be valid as long as the through space dipolar coupling PRE is the dominating relaxation mechanism for both T$_1$ and T$_2$. Furthermore, since this equation is independent of the strength of the coupling, it will hold for any nucleus whose relaxation is dominated by PRE and independent of the distance to the paramagnetic center (within the motional narrowing regime,\cite{pomr} which might not be the case for nearest neighbors with very large coupling strengths{, $\delta$, as the condition $\delta\tau_c\ll1$ might no longer be valid}).  In a sample dilute in paramagnetic species, each electron will be the source of relaxation for many nuclei. On the other hand, for a given nucleus relaxation will most likely be governed by a single paramagnetic center. In  the absence of efficient spin diffusion there will be a distribution of relaxation times, corresponding to the distribution of electron-nucleus distances. But also in this scenario, the ratio T$_1$/T$_2$  will be a single constant, as it is the same for every nucleus.
 While these assumptions should also be valid in the limit of fast spin diffusion, we do expect inconsistencies whenever spin diffusion affects longitudinal and transverse relaxation rates distinctly.
 
 Interestingly, computing the ratio of T$_1$/T$_2$ as a simple measure of correlation times is a common approach in NMR relaxometry of porous media.\cite{jpcc_113_6610_2009,cej_20_13009_2014,jacs_143_8249_2021,pccp_23_17752_2021} In this context, the correlation time is an indication of the liquid-solid adsorption properties.  \\

\begin{figure}
\begin{center}
\includegraphics[scale=1]{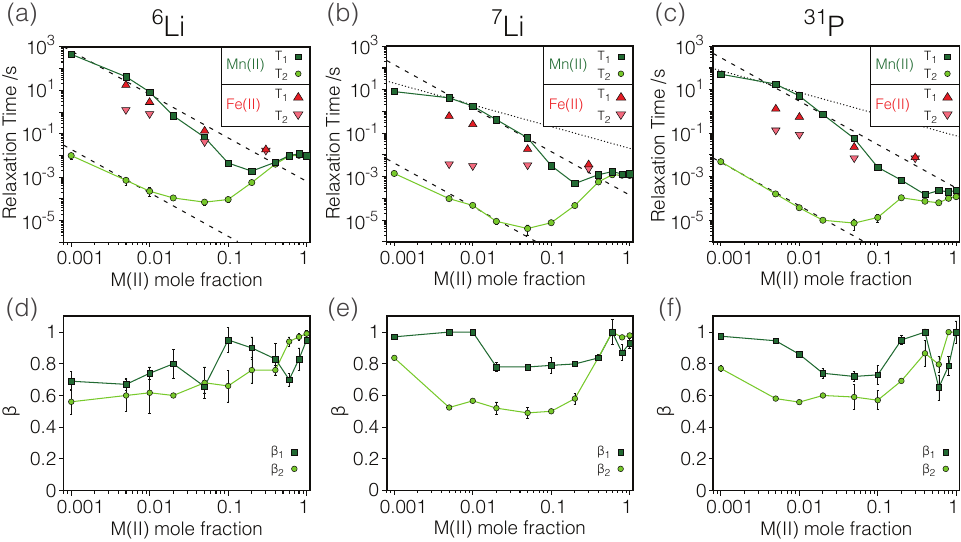}
\end{center}	
\caption{Longitudinal (dark green squares) and transverse (light green circles) relaxation times as well as corresponding stretch factors $\beta_{1,2}$ for $^6$Li (a) and (d), $^7$Li (b) and (e), and $^{31}$P (c) and (f) in LiMg$_{1-x}$Mn$_x$PO$_4$ for varying concentrations of Mn(II), obtained from equations~(\ref{eq:T1satrec}) and (\ref{eq:T2decay}). Measurements were done at room temperature and a spinning speed $\nu_R$ of 60~kHz. As guides for the eye, the dashed lines represent an inverse squared dependence of the relaxation times with the concentration of paramagnetic species (T$_{1,2}\propto [\text{Mn(II)}]^{-2}$), and the dotted line an inverse dependence (T$_{1,2}\propto [\text{Mn(II)}]^{-1}$). $^{31}$P measurements shown here for samples with high Mn(II) concentrations were done at carrier frequency offsets of 3000 ($x=0.4$ and 0.6) and 8000~ppm ($x=0.8$ and 1). Also shown in (a), (b) and (c) are the T$_1$ (dark red triangles) and T$_2$ (light red inverted triangles) relaxation times obtained in LiMg$_{1-x}$Fe$_x$PO$_4$.}
\label{fg:Relaxation}
\end{figure}

 Fig.~\ref{fg:Relaxation} shows the $^6$Li, $^7$Li and $^{31}$P T$_1$ and T$_2$ relaxation times in LiMg$_{1-x}$Mn$_x$PO$_4$ as a function of the Mn(II) concentration obtained by fitting the experimental data to equations~(\ref{eq:T1satrec}) and (\ref{eq:T2decay}). 
 Among the studied nuclei, $^6$Li has the lowest gyromagnetic ratio and natural abundance, therefore, due to the weak homonuclear dipolar couplings we do not expect spin diffusion to play a significant contribution in mediating relaxation. In the low concentration regime (which we loosely define as the region between $x=0.001$ and $x=0.01$) this is corroborated by the observed stretched factors, ranging between 0.5 and 0.7, indicative of a distribution of relaxation times, as well as by the inverse squared dependence of the relaxation times T$_1$ and T$_2$ with concentration (see dashed lines in Fig.~\ref{fg:Relaxation}a). The absence of significant $^6$Li spin diffusion is in line with our previous observations in similar materials.\cite{jpcl_11_5439_2020} 
 In this initial regime, despite the strong decay of relaxation times with increasing paramagnetic dopant content, the ratio T$_1$/T$_2$ remains approximately constant, as one would expect for a constant correlation time (equation~(\ref{eq:T1T2ratio})). Using equation~(\ref{eq:T1efromT1T2}) we estimate $\tau_{1e}$ as approximately 0.7~$\mu$s. This value is consistent with the trends observed in the direct measurements of T$_{1e}$ with EPR (Fig.~\ref{fg:EPRrelax}) at low temperatures, and is a good indication that in this case $\tau_{1e}\approx\text{T}_{1e}$, as we would expect in the absence of strong couplings among electron spins.

\begin{figure}
\begin{center}
\includegraphics[scale=1]{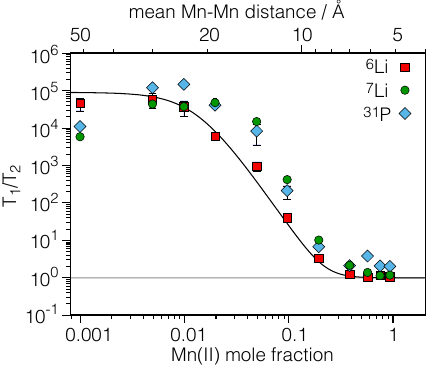}
\end{center}	
\caption{Ratio of T$_1$ over T$_2$ for $^6$Li, $^7$Li and $^{31}$P with variable concentration of paramagnetic Mn(II) dopant. Solid black line shows the behavior of $\text{T}_1/\text{T}_2$ for $\tau_{1e}\propto \left \langle r_{Mn-Mn}\right \rangle^6 $. Mean Mn-Mn distances were estimated using the nominal concentrations from the Wigner Seitz radii assuming a homogeneous distribution and a linear change in density.\cite{CIC_MIDNP_2021}}
\label{fg:T1overT2}
\end{figure}

Most interestingly, as we increase the Mn(II) content above $x=0.01$, we observe a strong decrease in the ratio T$_1$/T$_2$ (see also Fig.~\ref{fg:T1overT2}). From equation~(\ref{eq:T1efromT1T2}) we directly link this to a decrease of $\tau_{1e}$. These results suggest that we have reached a threshold of concentration of paramagnetic agents, after which, mean electron spin-spin interactions become significant and are directly responsible for a decrease in $\tau_{1e}$. At this point it is difficult to speculate whether this is a coherent or incoherent effect, however, we do want to highlight that the  $^6$Li  T$_1$/T$_2$ ratio decreases with mean Mn-Mn distance as $\tau_{1e}\propto\left \langle r_{Mn-Mn}\right \rangle^{6}$, as shown by the solid line in Fig.~\ref{fg:T1overT2}. A power of 6 dependence with mean distance indicates a second order contribution from the dipolar coupling and is most likely a consequence of relaxation mediated by through space dipolar couplings. This dependence is equivalent to $\tau_{1e}\propto [M]^{-2}$, where $[M]$ is the concentration of the paramagnetic metal, and has already been reported\cite{sjcp_23_231_1988} and discussed in terms of possible implications on the NMR relaxation times.\cite{joncs_188_54_1995} We are able to map the decay of $\tau_{1e}$ up to a concentration of approximately 0.2, after which the nuclear relaxation enters the extreme narrowing regime ($\omega_I\tau_{1e}<1$), where T$_1\approx\text{T}_2$ and thus the ratio is independent of $\tau_{1e}$. Additionally, above this point, the large concentration of Mn(II) species and short $\left\langle r_{Mn-Mn}\right\rangle$ will increase the relevance of Fermi contact and Heisenberg exchange interactions, as we have seen in the EPR spectra in Fig.~\ref{fg:EPRspectra}. In the presence of multiple relaxation mechanisms, the simple equation~(\ref{eq:T1T2ratio}) is not longer valid. Because of the fast drop of $\tau_{1e}$ both T$_1$ and, more pronounced, T$_2$ pass through a minimum at approximately $x=0.05$ after which the values increase with concentration. A direct consequence of this behavior is a reduction of the quenching radius leading to the observed increment in the $^6$Li signal intensity previously discussed.
 \\

 Next we turn our attention to the relaxation behavior of $^7$Li and $^{31}$P. In Fig.~\ref{fg:Relaxation} we do recognize the same general trends as for $^6$Li: very large initial ratio T$_1$/T$_2$, which stays relatively constant at low concentrations; followed by a steep decay of the ratio in the intermediate concentration regime including inversion of the concentration dependence of T$_1$ and T$_2$; finally leading to T$_2\approx\text{T}_1$ at large Mn(II) content.
 However, there are also some important differences between these nuclei and $^6$Li. The stronger homonuclear dipolar couplings are responsible for more efficient spin diffusion, leading to a $\beta_{1}$, the stretch factor of the longitudinal relaxation, approaching unity at the low concentration regime. Interestingly, in the intermediate Mn(II) concentration regime, between $x=0.01$ and 0.1 we observe a drop in $\beta_{1}$ to about 0.8 for both nuclei. We speculate that at this concentration a significant amount of the sample is relaxed directly by the dipolar coupling to the paramagnetic center due to efficient PRE. This effect might be further enhanced by a decay of the spin diffusion constant due to large line broadening in this concentration regime. This is likely also the reason for the strong deviation from a [Mn(II)]$^{-1}$ dependence of T$_1$ in this regime, as one would expect in case of efficient spin diffusion (compare dotted and dashed lines in Figs.~\ref{fg:Relaxation}b and \ref{fg:Relaxation}c).
 The transverse magnetization decay of  $^7$Li and $^{31}$P shows a remarkably prominent stretched behavior, with $\beta_{2}$ close to 0.5 over a large concentration range. Again, we attribute this to a very efficient transverse PRE mechanism, which outperforms homogenization of the spin coherence via spin diffusion.

 The distinct role of spin diffusion in longitudinal and transverse relaxation and the corresponding difference in $\beta$ impede an accurate determination of $\tau_{1e}$ from equations~(\ref{eq:T1T2ratio}) and (\ref{eq:T1efromT1T2}). Furthermore, at the lowest Mn(II) content of $x=0.001$, the $^7$Li and $^{31}$P T$_1$/T$_2$ ratio is significantly smaller than for $^6$Li and is probably indicative of the presence of an additional intrinsic relaxation mechanism. However, at concentrations above $x=0.01$, as $\beta_{1}$ and $\beta_{2}$ approach each other, estimations of $\tau_{1e}$ become more reliable, leading to very good agreement within the errors among all three nuclei (note that Fig.~\ref{fg:T1overT2} shows the ratio T$_1$/T$_2$ rather than $\tau_{1e}$, which would further require scaling by the nuclear Larmor frequency; $\tau_{1e}$ values are given in Tables~S1-S3).\\ 

 At very high Mn(II) content, relaxation mediated by the Fermi contact interaction becomes relevant and we observe differential relaxation across the lineshape for all nuclei (not shown). For the reported relaxation values we integrated over the entire excited region of the spectrum, this multi-exponential behavior is reflected in a drop of $\beta$ when the data is fitted with equations~(\ref{eq:T1satrec}) and (\ref{eq:T2decay}). As this occurs within the extreme narrowing regime, where we cannot access $\tau_{1e}$ from the ratio of T$_1$/T$_2$, we will not further elaborate on these differences.\\

 As an additional remark, we want to emphasize that  in principle, under complete absence of spin diffusion, the distribution of distances between paramagnetic species and nuclei should lead to an analogous distribution of longitudinal and transverse relaxation times. Small variations in the obtained $\beta_{1,2}$ values are indicative of a deviation from the ideal model. It is therefore important to note that $\tau_{1e}$ values obtained from the ratio of T$_1$/T$_2$ will most likely have a larger uncertainty  than suggested from the error propagation of the individual measurements. Nonetheless, the reported results strongly suggest that we are able to give good estimates of the size of $\tau_{1e}$, at least within one order of magnitude.\\

Finally, we also measured the relaxation times of $^6$Li, $^7$Li and $^{31}$P in the Fe doped olivine for various Fe(II) concentrations, these are also shown in Fig.~\ref{fg:Relaxation}. Due to the very short T$_{1e}$ of Fe(II) this paramagnetic species is very inefficient in causing nuclear transverse relaxation. Consequently, the ratio of T$_1$/T$_2$ is much smaller than in the Mn(II) doped olivine. Deviation from unity at low Fe(II) concentrations, most accentuated for $^7$Li, are attributed to intrinsic coherence loss, which is independent of the concentration of paramagnetic species. Furthermore, unlike for the Mn(II) containing samples, here the concentration dependency of T$_1$ follows a constant trend, indicating that contributions from electron-electron dipolar couplings do not significantly add to the fast relaxation rate of Fe(II).

\subsubsection{DNP}

\begin{figure}
\begin{center}
\includegraphics[scale=1]{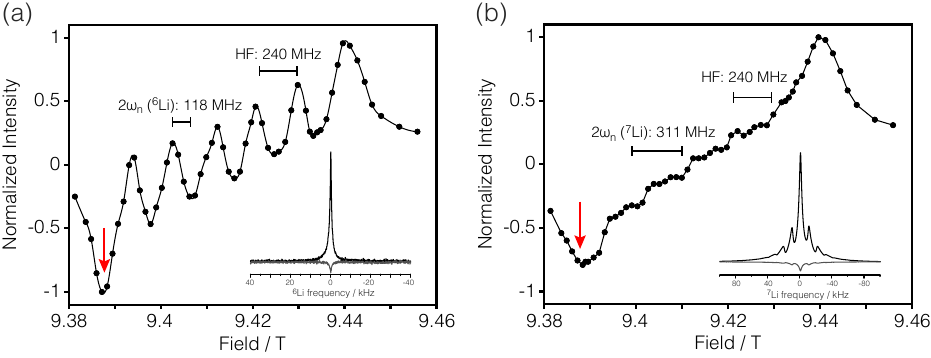}
\end{center}	
\caption{MAS DNP Field sweeps of $^6$Li (a) and $^7$Li (b) in LiMg$_{1-x}$Mn$_x$PO$_4$  with $x=0.005$ obtained at 100 K, with a spinning speed of $\nu_R$ of 9.7~kHz and using a recycle delay of 30 and 5~s, respectively. The insets show the spectra obtained with and without microwave irradiation, and using recycle delays of 256 and 4096~s for $^6$Li and $^7$Li, respectively, obtained enhancements $\epsilon_{\text{ON/OFF}}$ were 11 and 12.1, respectively. Red arrows indicate the magnetic field  at which the spectra shown in insets were acquired. }
\label{fg:DNPSweep}
\end{figure}

 CW DNP requires long electron relaxation times for most efficient saturation of the irradiated transition. This is particularly critical for the solid effect mechanism, as the relevant double and zero quantum transitions are formally forbidden, thus the effective nutation frequency is scaled down by the nuclear Larmor frequency,\cite{emagres_8_295_2019} making saturation at high magnetic fields a very challenging task. As the solid effect is the most common DNP mechanism for metal ions in inorganic solids, we expect that the obtained $\tau_{1e}$ values should correlate with the DNP efficiencies in these materials. {Here we verify qualitatively this behavior, while a more in-depth discussion regarding the relation between electron relaxation rates and DNP enhancements will be presented in the next section.} 
The MAS DNP field sweep profiles of $^6$Li and $^7$Li  in LiMg$_{1-x}$Mn$_x$PO$_4$  with $x=0.005$ are shown in Fig.~\ref{fg:DNPSweep}. 
In the $^6$Li sweep the characteristic lineshape arising from the splitting of the Mn electron spin signal due to the hyperfine coupling to the $^{55}$Mn nucleus is observed.\cite{pccp_18_27205_2016,jacs_141_451_2019} Partial cancellation of positive and negative enhancement lobes does not allow to resolve the splitting in the $^7$Li sweep. At this concentration we obtain signal enhancements of 11 and 12.1 for $^6$Li and $^7$Li, respectively. At lower concentration ($x=0.001$) we obtain similar enhancement for $^6$Li  ($\epsilon_{ON/OFF}=9.1$) as one would expect from the ratio of T$_1$/T$_2$, which does not vary significantly among these concentrations, and assuming that the observed concentration dependence does not vary significantly with temperature, since relaxation measurements were performed at room temperature while DNP at 100~K. On the other hand, the $^7$Li enhancement at $x=0.001$ is only about 2.5. We attribute this to the presence of non-PRE relaxation mechanisms.\cite{jpcl_11_5439_2020} As we go to larger concentrations we observe the expected decay in enhancement factor accompanying  the decay in the T$_1$/T$_2$ ratio ($^7$Li  $\epsilon_{ON/OFF}=3.3$ at $x=0.02$, while no enhancement was observed for $x=0.2$; all measured values are summarized in Table~S4).

\subsection{Variable temperature}

\begin{figure}
\begin{center}
\includegraphics[scale=1]{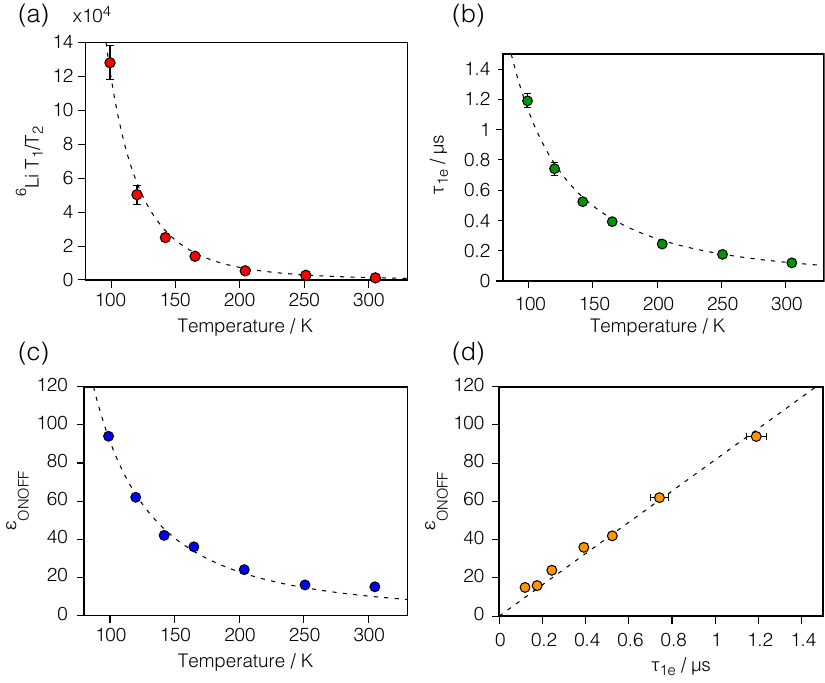}
\end{center}	
\caption{Variable temperature $^6$Li relaxation times and DNP enhancements in Li$_4$Ti$_5$O$_{12}$ doped with Fe(III) with a mole fraction of $x=0.01$. Longitudinal and transverse relaxation times, T$_{1}$ and T$_{2}$, were obtained from fit of the experimental data to equations~(\ref{eq:T1satrec}) and (\ref{eq:T2decay}), with a constant $\beta_{1,2}=0.6$. (a) Ratio of T$_{1}$ and T$_{2}$ as a function of temperature.  (b) Correlation time of the fluctuations of the longitudinal component of the electronic magnetic moment, $\tau_{1e}$, estimated according to equation~(\ref{eq:T1efromT1T2}) as a function of the temperature. (c) DNP enhancement $\epsilon_{\text{ON/OFF}}$ obtained from the maximum intensity of the signal as a function of the temperature. (d) $\epsilon_{\text{ON/OFF}}$ against $\tau_{1e}$ obtained for variable temperatures. All measurements were done at 10~kHz MAS. The dashed lines show best fits to an equation of the type $f(x)=mx^n$, with a fixed value of $n$. The following parameters were obtained: (a) $n=-4$, $m=1.19\pm0.03\times10^{13}$; (b) $n=-2$, $m=82\pm2$; (c) $n=-2$, $m=920000\pm20000$; (d) $n=1$,  $m=11200\pm200$.}
\label{fg:FeLTO_summary}
\end{figure}

Finally, in an effort to confirm the correlation between the ratio of T$_{1}$/T$_{2}$ and the DNP enhancement we studied the temperature dependence of both parameters in Fe(III) doped  Li$_4$Ti$_5$O$_{12}$. The large $^6$Li DNP enhancement, $\epsilon_{\text{ON/OFF}}$, at 100~K of about 100\cite{jpcc_124_7082_2020} enabled measuring sizable enhancements over a broad temperature range, up to room temperature. Further, we have shown in a previous study that the polarization enhancement of $^6$Li in this sample is governed by direct polarization rather than spin diffusion,\cite{jpcl_11_5439_2020} making this an ideal sample for this case study.
 Fig.~\ref{fg:FeLTO_summary}a shows the ratio of the measured $^6$Li T$_{1}$ and T$_{2}$ relaxation times as a function of the temperature. Using equation~(\ref{eq:T1efromT1T2}) we obtain the correlation time $\tau_{1e}$ (Fig.~\ref{fg:FeLTO_summary}b). At this low concentration of paramagnetic species, $\tau_{1e}$ is likely a good estimate of the electronic relaxation time T$_{1e}$. The value of 1.2~$\mu\text{s}$ obtained at 100~K and 9.4~T is in reasonable agreement with previously reported T$_{1e}$ of 1.5~$\mu\text{s}$ for the same sample at 50~K and 1.2~T.\cite{jpcc_124_7082_2020} We observe that $\tau_{1e}$ decreases with T$^{-2}$, this is the expected behavior of Raman activated electron spin relaxation, which further indicates that in this case $\tau_{1e}$ is a good measure of T$_{1e}$.
 The DNP enhancement as a function of the temperature is shown in Fig.~\ref{fg:FeLTO_summary}c, as can be seen, it also decreases with a T$^{-2}$ dependence (dashed line). Consequently, we find a linear relation between $\epsilon_{\text{ON/OFF}}$ and $\tau_{1e}$ (Fig.~\ref{fg:FeLTO_summary}d) within the studied temperature range. \\

The solid effect DNP enhancement via direct polarization can be obtained from the saturation efficiency:\cite{jpcl_11_5439_2020}
\begin{equation}
P_{n}/P_e=\pm\frac{P_{e,eq}-\Delta p_{DQ/ZQ}}{P_{e,eq}},
\label{eq:Enhancement}
 \end{equation}
here P$_n$ and P$_e$ are the nuclear and electron polarization and $\Delta p_{DQ/ZQ}$ is the population difference of the states involving the double or zero quantum transition:
\begin{equation}
\Delta p_{DQ/ZQ}=
\frac{
\Delta p_{DQ/ZQ,eq}}
{1+\omega_{1,DQ/ZQ}^2R_{2e}^{-1}\left(2R_{1,DQ/ZQ}+2R_{1n}\right)^{-1}},
\label{eq:DQsaturation}
 \end{equation}
where $\omega_{1,DQ/ZQ}$ is the effective nutation frequency of this transition, and R$_{1,DQ/ZQ}$ its longitudinal relaxation rate (R$_{1,2}=\frac{1}{\text{T}_{1,2}}$) which is proportional to R$_{1e}$.\cite{jmr_207_176_2010} When PRE is the main source of relaxation we can use equation~(\ref{eq:T1PRE}) to obtain the nuclear relaxation rate R$_{1n}$, which is also proportional to R$_{1e}$. Thus: 
\begin{equation}
\Delta p_{DQ/ZQ}\propto\frac{1}{1+\omega_1^2T_{1e}T_{2e}}.
\label{eq:DQpropto}
 \end{equation}
 
\begin{figure}
\begin{center}
\includegraphics[scale=1]{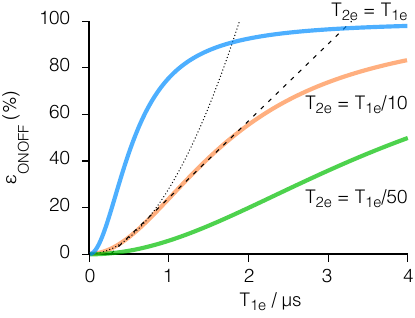}
\end{center}	
\caption{Calculated steady-state DNP enhancement $\epsilon_{\text{ON/OFF}}$ as a function of the electronic relaxation times with various ratios T$_{2e}/$T$_{1e}$ following equations~(\ref{eq:Enhancement}) and (\ref{eq:DQsaturation}) assuming a single electron spin, coupled to a single nucleus with the gyromagnetic ratio of $^6$Li in a 9.4~T magnetic field at 100 K under 0.35 MHz microwave irradiation\cite{pccp_21_2166_2019} (multiplied by factor of 3 to account for the effective central transition nutation frequency of a spin $S=5/2$) and on-resonance with the double quantum transition. As guide for the eye, a linear (dashed line) and squared (dotted line) function are included. }
\label{fg:Simulation}
\end{figure}

The behavior of the enhancement factor with electron relaxation time predicted by these equations is shown in Fig.~\ref{fg:Simulation}. At low enhancement factors the curve follows a squared dependence, assuming that T$_{2e}\propto\text{T}_{1e}$. However, over a large region around the saddle point of the curve the enhancement follows a nearly linear behavior. This is in agreement with the observed linear trend in the experimental results. We note that equation~(\ref{eq:DQpropto}) predicts the same dependency of the enhancement from the nutation frequency $\omega_1$ and the electron relaxation time (again, assuming  T$_{2e}\propto\text{T}_{1e}$).\cite{jmr_207_176_2010} 

The maximum theoretical enhancement for $^6$Li from DNP is approximately 4470, however, as most likely only the central transition contributes to the enhancement significantly, this number gets scaled down by factor 3.\cite{jacs_133_5648_2011} Consequently, the enhancement factor 100 represents about 7\% of the theoretical limit. In addition, it is important to note that homogeneous and inhomogeneous broadening of the EPR line shape will also play a critical role in the observed DNP enhancements. Due to partial cancellation of the positive and negative enhancements a further decrease of the maximum theoretical value of $\epsilon_{\text{ON/OFF}}$ is expected. In this regard, the simulations shown in Fig.~\ref{fg:Simulation} show a remarkably good agreement with the experimental results when assuming a ratio T$_{2e}=\text{T}_{1e}/10$, which is in line with the values in LiMg$_{1-x}$Mn$_x$PO$_4$ at the highest measured temperatures (Fig.~\ref{fg:EPRrelax}).

\section{Conclusions}

 In this paper we studied how the amount of paramagnetic species in inorganic solids affects the NMR relaxation behavior. Synthesis of the solid solution LiMg$_{1-x}$Mn$_x$PO$_4$ enabled us to cover the entire range of concentrations, from extremely dilute, where paramagnetic Mn(II) is only a minor dopant, to very high, where it is one of the main components of the structure.
 Through a careful analysis of the EPR and NMR properties in this system, we found that changes in the nuclear relaxation times T$_1$ and T$_2$ and, more specifically, their ratio, can be used to map changes in the correlation time of the fluctuations of the electronic magnetic moment, $\tau_{1e}$. Further, we discuss large differences in NMR spectra due to variable amount of paramagnetic species and  are able to understand the observed broadening and quenching factors based on changes in  $\tau_{1e}$.

 In addition, this approach enables rationalizing and facilitates a priori prediction of DNP enhancements in metal ions based DNP simply from the nuclear relaxation times T$_1$ and T$_2$, obtained at the conditions of interest. This latter point is a clear advantage compared to characterization of the electron spin via EPR, which often, due to instrumental limitations, cannot be performed at the same field or temperature  as the high field DNP experiments, and therefore require additional assumptions to predict the relevant properties. An important prerequisite for this approach is that both T$_1$ and T$_2$ are dominated by the PRE from the same paramagnetic center which will act as polarizing agent upon microwave irradiation. Most notably, we have found experimentally a linear dependence of the DNP enhancement and $\tau_{1e}$, determined from the ratio T$_1$/T$_2$, in Fe(III) doped Li$_4$Ti$_5$O$_{12}$ over a temperature range between 100 and 300~K.

\section{Acknowledgements}
We thank Dr.~Brijith Thomas for assistance with XRD measurements, Guy Reuveni for synthesizing Li$_4$Ti$_5$O$_{12}$ and Prof.~Philip Grandinetti for helpful discussions.  This research was funded by the European Research Council (MIDNP, grant no. 803024). The work was made possible in part by the historic generosity of the Harold Perlman family.

\bibliography{Biblio-Bibtex} 

\end{document}